Graphical Abstract: POLYMER-STABLE MAGNESIUM NANOCOMPOSITES PREPARED BY LASER ABLATION FOR EFFICIENT HYDROGEN STORAGE

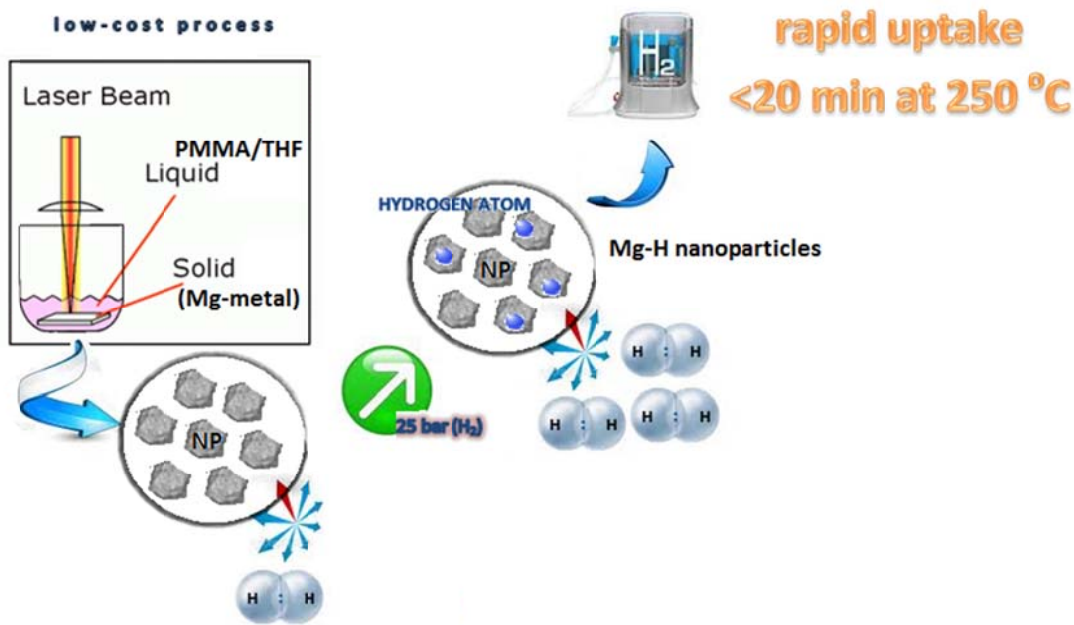

Laser ablation shows great potential as a scalable method for the production of metal hydrides due to its capability of fabricating single phase, fine nanoparticles with a maximum dispersion while being a relatively low-cost process. The encapsulation of the Mg nanoparticles in a polymer matrix allowed to achieve a rapid uptake (<20 min at 250 oC) of hydrogen with a high capacity (6 wt. % in Mg, 5.5 wt. % overall).

Highlights

- Laser ablation for high efficient polymer composite air stable Mg-nanoparticles.
- Mg nanoparticles are well dispersed and single phase below 5 nm.
- Hydrogen absorption-desorption kinetics prove ultrafast reactions.
- The PCM Mg-nanoparticles are fully re-chargeable.
- Laser ablation is a promising technique in hydrogen technologies.

# POLYMER-STABLE MAGNESIUM NANOCOMPOSITES PREPARED BY LASER ABLATION FOR EFFICIENT HYDROGEN STORAGE

S.S. Makridis[1,2,*], E. Gkanas[1,2], G. Panagakos[1], E.S. Kikkinides[2], A.K. Stubos[1], P. Wagener[3] and S. Barcikowski[3]

[1]*Environmental Technology Laboratory, Institute of Nuclear Technology and Radiation Protection, NCSR 'Demokritos', Agia Paraskevi, Athens, 15310, Greece*

[2]*Department of Mechanical Engineering, University of Western Macedonia, Bakola & Sialvera Street, Kozani, 50100, Greece*

[3]*Technical Chemistry I and Center for Nanointegration Duisburg-Essen (CENIDE), University of Duisburg, Universitaetsstrasse 7, D-45141 Essen, Germany*

*correspondence: sofmak@ipta.demokritos.gr

**Abstract**

Hydrogen is a promising alternative energy carrier that can potentially facilitate the transition from fossil fuels to sources of clean energy because of its prominent advantages such as high energy density (142 MJ kg-[1]), great variety of potential sources (for example water, biomass, organic matter), and low environmental impact (water is the sole combustion product). However, due to its light weight, the efficient storage of hydrogen is still an issue investigated intensely. Various solid media have been considered in that respect among which magnesium hydride stands out as a candidate offering distinct advantages.

Recent theoretical work indicates that $MgH_2$ becomes less thermodynamically stable as particle diameter decreases below 2 nm. Our DFT (density functional theory) modeling studies have shown that the smallest enthalpy change, corresponding to 2 unit-cell thickness (1.6 Å Mg/3.0Å $MgH_2$) of the film, is 57.7 kJ/molMg. This enthalpy change is over 10 kJ/molMg smaller than that of the bulk. It is important to note that the range of enthalpy change for systems that are suitable for mobile storage applications is 15–24 kJ/molH at 298 K.

The important key for the development of air-stable Mg-nanocrystals is the use of PMMA (polymethylmethacrylate) as an encapsulation agent. In our work we use laser ablation, a non-electrochemical method, for producing well dispersed nanoparticles without the presence of any long range aggregation. The observed improved hydrogenation characteristics of the polymer-stable

Mg-nanoparticles are associated to the preparation procedure and in any case the polymer-laser-ablation is a new approach for the production of air-protected and inexpensive Mg-nanoparticles.

I.      Introduction

The continuous growth of world population and the intense economic expansion of developing countries are among the major causes of the increasing demand for energy and the alarming and continuous release of greenhouse gases. Among several scenarios, hydrogen is the most promising energy carrier to satisfy the required conditions for the ideal fuel. It is the cleanest fuel and has a heating value three times higher than petroleum. While it seems to be the ideal means of transport and conversion of energy for mobile and stationary applications, a major problem is the storage of hydrogen which presents several issues mainly related to safety and amount of stored hydrogen [1].

In recent years a lot of research has been done in materials for hydrogen storage. Magnesium-based alloys have attracted much attention due to high hydrogen capacity and low cost. It is reported that pure Mg can store up to 7.6 wt% [1-3]. For pure Mg, the hydrogenation enthalpy is around −74.7 kJ/mol H2 and its activation energy is evaluated to be 86 kJ/mol $H_2$ [4,5].

Despite the relatively high capacity, there are certain important disadvantages of using Mg alloys such as slow kinetics, high operation temperatures and high reactivity with oxygen [6-10] which constitute significant obstacles for practical on-board applications.

Considerable research has been conducted on magnesium metals to synthesize new high performance materials and develop more efficient techniques. These studies are mainly focused on the a) element substitution [11] b) new production methods by using for example different hydrogen pressures [12,13] c) preparation of composite materials through the addition of dopants in order to improve microstructure/microchemistry [14] and d) annealing [15] in order to improve hydrogen storage characteristics [16].

It is known that preparation and synthesis methods are important factors which can influence the characteristics of the samples. Mechanical alloying (MA) and ball milling (BM) are widely used for preparation of Mg materials [8,11, 17-19]. For Powder Metallurgy (PM) an important step to synthesize these compounds is sintering which improves the bonding between the powders and minimizes the porosity [20]. Unfortunately, all these methods require long times and can cause contaminations to the alloys even under protective atmosphere.

Another known technique is the hydriding combustion synthesis (HCS) which can produce much purer samples but needs a stable temperature for several hours [12, 21-23].

In order to eliminate the problems caused by conventional methods, a new rapid heating technique should be used. Microwave heating is a technology that is mostly used for ceramics, carbides and ferrites and has not been applied for metals due to the fact that metals reflect microwaves. Gupta et al. [24] reported for the first time that Mg alloys can be synthesized by hybrid microwave heating [25-27]. Li et al. [28] prepared $Mg_2Ni$ alloys using microwave-assisted activation synthesis (MAAS) and showed that these materials can absorb 3.2 %wt of $H_2$ in only 50 s at 523 K under 3MPa $H_2$. Wong et al. [29] studied the effect of microwaves on the structural and microstructural characteristics of Mg-based compounds. His team revealed that there were no defects and the surface was smooth and free of radial and circumferential cracks. They also showed that there was an increase in hardness with the addition of nanometer – scale reinforcements. In short, heating by microwaves has certain advantages over the conventional methods such as a) reduction of processing time, b) uniform heating, c) improved properties and d) environmental friendliness [30].

Recently, nanomaterials have attracted a great interest because of their unique characteristics, different from bulk materials. Therefore, a wide range of synthetic approaches regarding the preparation of metal nanoparticles in various matrices, including reduction method, sol-gel process, solvent evaporation of hydrophobic colloids have been reported [31-33]. Laser ablation which is usually applied to in-situ elemental analysis [34], forming thin film (PLD: pulsed laser deposition) [35] has been also used to prepare nanoparticles [36-43]. In particular, the temperature and pressure of the plume induced by pulsed laser irradiation onto the metal target surface in liquid are very high [44, 45] compared to their values in vacuum or atmosphere because of the confinement effect. Since there are ablated particles in this plume, crystalline nanoparticles can be obtained without any heat treatments [46]. The laser ablation method offers a great advantage regarding the production of nanoparicles while aggregation and dispersion can be controlled by using surfactant or any appropriate liquid [47-49]. So, by this technique pure nanoparticles can be obtained and be captured in principle in any liquid. As ablated particles go through the plume induced by laser ablation in liquid in which the temperature and pressure are very high, nanoparticles with new optical, electrical and mechanical properties can be expected to be fabricated.

The consideration of new Mg-type of materials or comsposites going to nanoscale fabrication for more efficient hydrogen storage technologies has extensively been studied by other groups [50-59]. By decreasing theoretically the particle size diameter below 2 nm, the nanoparticles of magnesium hydride have less thermodynamic stability [60, 61]. Density functional theory studies have shown that the smallest enthalpy change is 57.7 kJ/molMg and corresponds to a two unit-cell thickness (1.6 Å Mg/3.0Å $MgH_2$) of the thin film. It is really important that this enthalpy change is over 10 kJ/molMg smaller than that of the pure bulk magnesium. At room temperature, the desired and targeted range of enthalpy change for mobile storage systems technology is 15–24 kJ/molH [62].

According to the published information on hydrogen storage technology, there is a need of air-stable materials and especially in magnesium hydrides with relatively nanosize grains, the key for the development of the air-stable nanocrystals [63] is the PMMA (polymethylmethacrylate) polymer. In order to overcome chemical or electrochemical ways of producing air-stable nanoparticles we used the laser ablation technique. It is suggested as one of the most reliable techniques for having well dispersed nanoparticles with no long range aggregations.

II. Experimental

Laser ablation of solid targets into liquids is a quite new alternative physical method for fabrication of nanoparticles. The metal target was placed on the bottom of a glass cuvette filled with 1.5 mL of pure deionized water and in polymer matrices. The rod was irradiated with the fundamental (1,064 nm) of a Nd:YAG laser (Quanta-Ray GCR-190, Spectra Physics) operating at 10 Hz on a rotating base with speed 4.5 degrees/s. The laser beam was focused 1 mm below the surface target, and the irradiation time was kept constant (30 min). Upon irradiation, the solution gradually turned in different experiments from light red to wine red. The absorption spectra of the colloidal solutions were measured immediately after fabrication by a Cary UV–Visible spectrophotometer. The above procedure in polymer matrix may reveal very well dispersed nanoparticles below 5 nm.

X-ray diffraction analysis of the alloys was carried out on the powders at room temperature by using Cu-K$_\alpha$ radiation, in a SIEMENS D500 X-Ray diffractometer. Rietveld analysis has been performed on the XRD patterns with the use of the RIETICA software. Nanoanalysis has been performed by using a HR-TEM operating at 200 kV and equipped with a spherical aberration corrector in the objective lens, to ensure a point resolution of 1.2 A. The SCION image software has been used for the particle size analysis. The hydrogenation/dehydrogenation kinetics and cycle stability of the sample have been studied, using a Magnetic Suspension Balance (Rubotherm). In this equipment, hydrogen desorption and re-absorption, can be investigated at constant hydrogen pressures in the range from 1 to 20 MPa (flow-through mode).

III. Results and discussion

The nanoMg/PMMA composites were synthesized at room temperature from a homogeneous tetrahydrofuran (THF) solution containing the gas-selective polymer poly(methyl methacrylate) (PMMA). This polymer has been used in the sol-gel type technique elsewhere [63] and air stable high quality nanoparticles have been produced. Laser ablation has been used to produce rapidly and more efficiently well dispersed nanoparticles for upscaling purposes in the hydrogen storage technology. We target on the development of low cost and air stable high surface metal nanodispersions in polymer matrix.

The principle of the laser-based synthesis of nanoparticles in liquids is illustrated in figure 1. Here a pulsed laser beam is focused on a target in a solvent. After absorption of the laser pulse energy, the target material is vaporized and condenses in the solvent thus forming nanoparticles. The use of ultra-short pulses enables application of volatile organic solvents or monomers. In general, every combination of target material and dispersion phase is possible.

By varying the manufacturing process numerous material combinations can be tested in short time. The laser process can be called "rapid nanomaterial prototyping" or regarding the composite synthetics as "rapid nanocomposite manufacturing", both because of the nearly unlimited material variety and because it is easy to adapt the parameters [64]. Either metallic or ceramic nanoparticles can be generated in aqueous or organic solvents [65].

*Fig. 1. Laser-generated Nanoparticles for Nanocomposites (solution turns darker with the experiment, which implies the increase in nanoparticle concentration)*

The X-ray diffraction pattern of the as ablated nanoparticles in the PMMA matrix, as shown in figure 2, shows characteristic reflexes of a single-phase hexagonal magnesium. This is attributed to a very high quality ablation in the composite matrix so that the nanoparticles do not react with the air. As an inset, it can be seen that there are no reflexes that belong to Mg-O or Mg-OH phases that deteriorate the hydrogenation capacity (it is well known the negative effect of oxidations on the metal hydride efficiency).

Fig. 2. XRD pattern of Mg-nanoparticles and as inset are shown the PCPDF cards of the Mg, MgO and Mg(OH)$_2$. Only Mg (hkl) planes have been traced and indexed. Cards: Hexagonal Mg (solid black line, JCPDS 04-0770), cubic MgO (dashed black line, JCPDS 89-7746), and hexagonal Mg(OH)$_2$ (pale grey line, JCPDS 07-0239). Crystallographic characteristics: a=3.212 (2) Å, c=5.218 (4) Å

Fig. 3. TEM (up), HR-TEM micrograph and particle size estimation (down) of laser ablated nanoMg/PMMA composite

As revealed from the analysis of the XRD line profile broadening of each reflex, the grain size can be estimated. By using the well-known Scherrer's formula we estimated that the grain-particle size of the nanoparticles is below 7 nm. This means that the particles in our analysis present almost half the size in comparison to the chemically produced air-stable composite Mg nanoparticles, as it reveals from an already published work [63]. In our case, as shown in figure 3, the estimated particle size from the TEM analysis is very close to the experimental value. The results of the line profile analysis of the reflexes showed that the particles are much better dispersed and this should be associated to the fact that the absorption/desorption analysis is faster and more efficient in the laser ablated nanoparticles, as shown in figure 4, compared to other published results [63].

The fact that the particle size estimation from the TEM micrograph analysis by using the SCION image software and the XRD grain size calculations are very close suggests that the particle dispersion is much better than the sol-gel method discussed in the recently published literature [63 & supplementary data]. As shown in figure 4, the kinetics and hydrogen capacity for the laser ablated Mg nanoparticles of this work are very comparable to those of reference [63] but the reversible hydrogen content is considerably better (~96 % of the absorbed hydrogen amount was desorbed at 250 $^o$C in less than 20 min) since full re-chargeability has been found in our case under similar conditions. Grain-particle size distribution in the polymer matrix seems to play a mandatory role in the hydrogen reversibility and kinetics, as expected. After activation procedure, in all three charges we obtain fully saturated nanostructured materials since no difference in the curves was found in the hydrogenation/dehydrogenation procedure. A slight difference in the desorbed amount of hydrogen under both 1 bar and $10^{-2}$ bar (vacuum) was observed after each discharging. The absorbed amount is the same at the highest pressure of 25 bar of hydrogen. The reversibility is almost the same in all three curves and the slight difference between the three discharges could be attributed to the Mg to Mg-H ratio in the polymer matrix or to the hydrogen molecules remaining as adsorbed amount at the boundaries of Mg to polymer. Those assumptions are according to the fact that realistically, at the nanoscale, hydrogen may be adsorbed also on the surface of the particle (or film).

*Fig. 4. Hydrogenation/dehydrogenation process in the magnetic suspension balance equipment: Mg composite nanoparticles in PMMA*

IV. Conclusions

Novel Mg nanoparticles have been synthesized in a polymer matrix by using laser ablation, an ultrafast and highly efficient method. The latter is found to be a powerful method to produce very fine, well dispersed, single phase nanoparticles. It has been considered as a potential efficient technique for upscaling the production of high surface metallic particles in the hydrogen storage technology. By encapsulation in a polymer matrix, the Mg nanoparticles exhibit more rapid, compared to other Mg-pure types, uptake of hydrogen (<20 min at 250 $^o$C) with a high capacity (6 wt. % in Mg, 5.5 wt. % overall). Compared to the sol-gel obtained nanoparticles published elsewhere [63], the laser ablated nanoparticles have excellent reversibility under vacuum and at 250 $^{o}$C, a relatively low temperature with regard to the necessary ~330 $^{o}$C for Mg-bulk materials.

In future work other types of intermetallic hydrides or polymer-types could be considered for the development of novel materials for hydrogen storage purposes.

## Acknowledgements

This work was partially supported by the ATLAS-H2 IAPP European Project (Grant Agreement 251562)

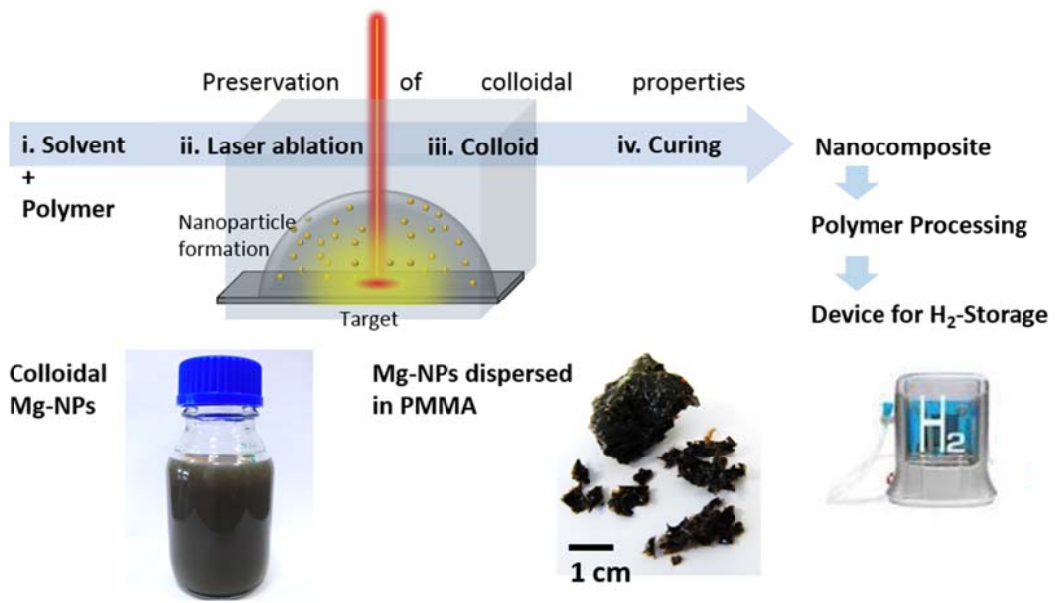

Fig. 1. Laser-generated Nanoparticles for Nanocomposites

Makridis et al. **POLYMER-STABLE MAGNESIUM NANOCOMPOSITES PREPARED BY LASER ABLATION FOR EFFICIENT HYDROGEN STORAGE**

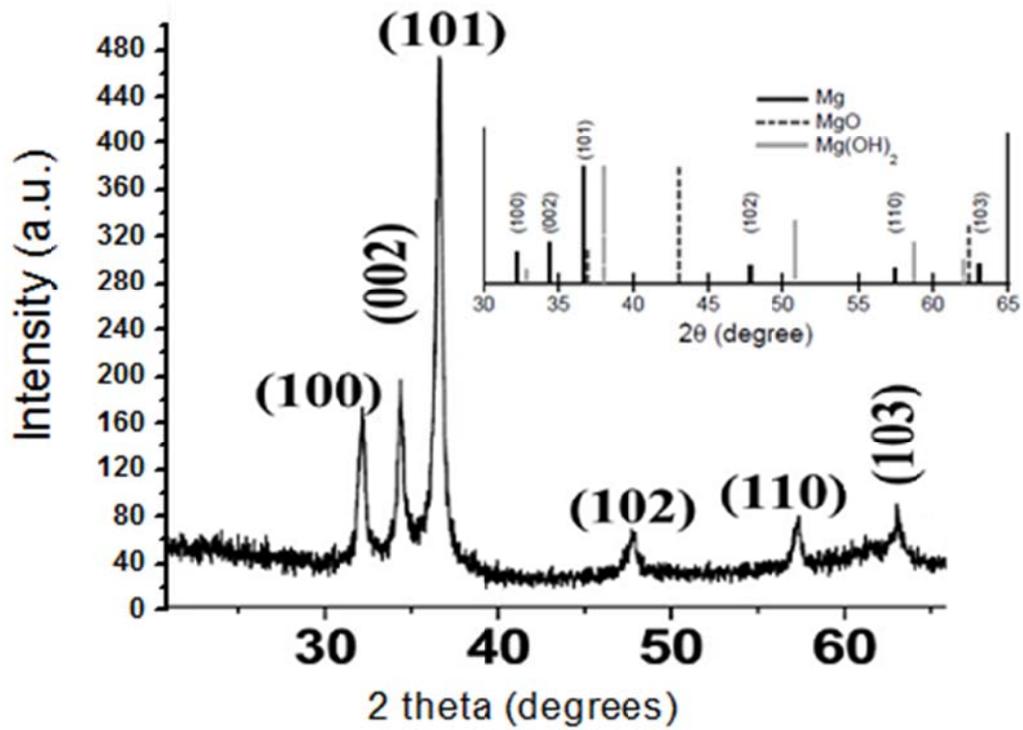

Fig. 2. XRD pattern of Mg-nanoparticles and as inset are shown the PCPDF cards of the Mg, MgO and Mg(OH)$_2$. Only Mg (hkl) planes have been traced and indexed. Cards: Hexagonal Mg (solid black line, JCPDS 04-0770), cubic MgO (dashed black line, JCPDS 89-7746), and hexagonal Mg(OH)$_2$ (pale grey line, JCPDS 07-0239). Crystallographic characteristics after Rietveld analysis: a=3.212 (2) Å, c=5.218 (4) Å

Makridis et al. **POLYMER-STABLE MAGNESIUM NANOCOMPOSITES PREPARED BY LASER ABLATION FOR EFFICIENT HYDROGEN STORAGE**

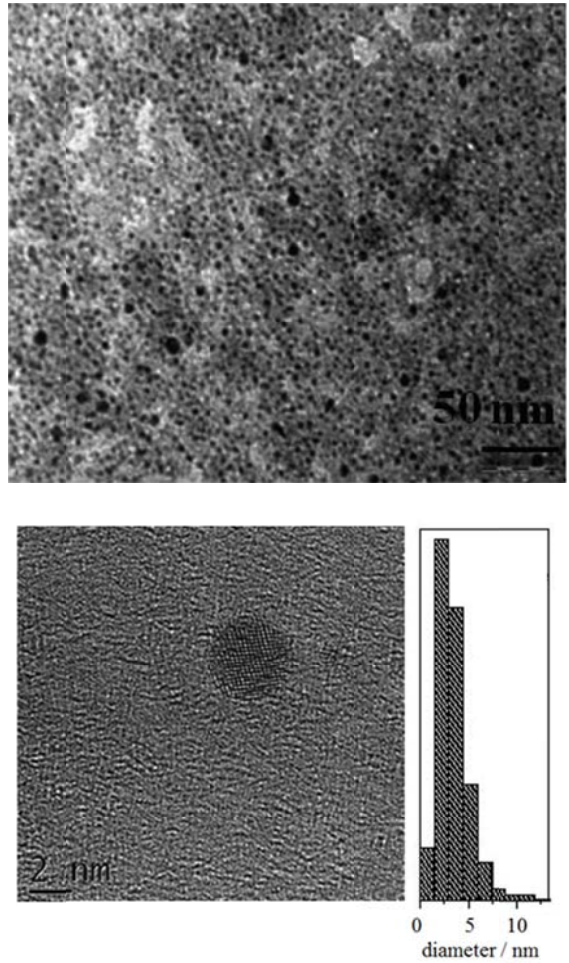

Fig. 3. TEM (up), HR-TEM micrograph and particle size estimation (down) of laser ablated nanoMg/PMMA composite

Makridis et al. **POLYMER-STABLE MAGNESIUM NANOCOMPOSITES PREPARED BY LASER ABLATION FOR EFFICIENT HYDROGEN STORAGE**

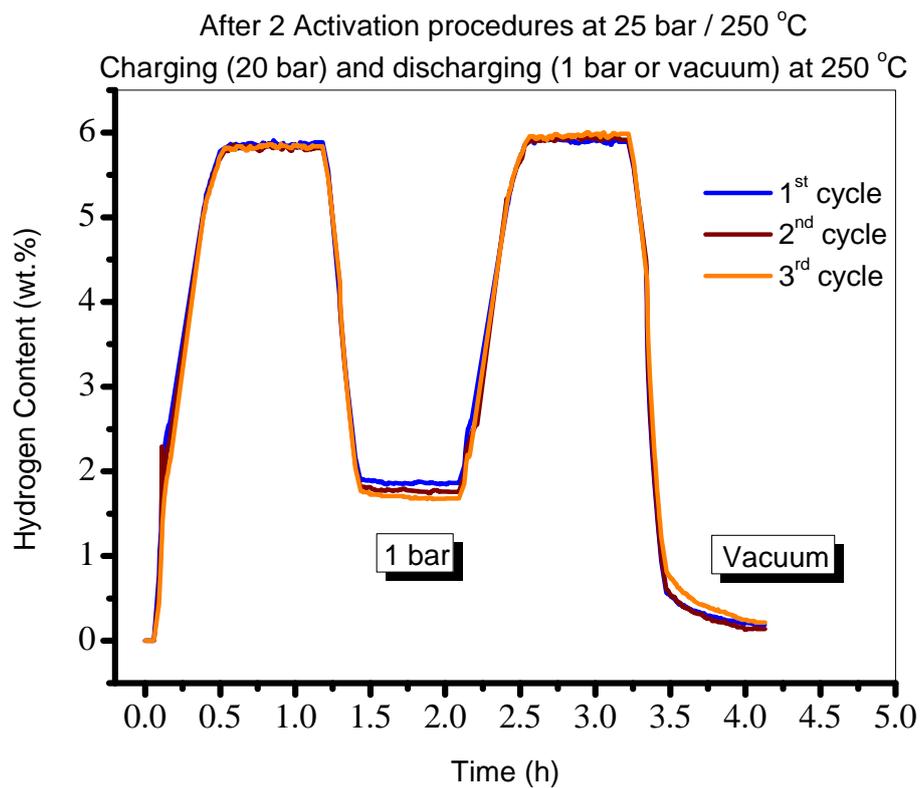

Fig. 4. Hydrogenation/dehydrogenation process in the magnetic suspension balance equipment: Mg composite nanoparticles in PMMA

Makridis et al. **POLYMER-STABLE MAGNESIUM NANOCOMPOSITES PREPARED BY LASER ABLATION FOR EFFICIENT HYDROGEN STORAGE**